\documentclass[12pt,preprint]{aastex}

\usepackage{graphics,graphicx}

\usepackage[usenames]{color}

\begin{document}

\title{The Discovery of the First ``Changing Look'' Quasar: New Insights into the Physics \& Phenomenology of AGN}

\author{Stephanie M. LaMassa$^{1}$, Sabrina Cales$^{2,1}$, Edward C. Moran$^3$, Adam D. Myers$^4$, Gordon T. Richards$^5$, Michael Eracleous$^6$, Timothy M. Heckman$^7$, Luigi Gallo$^8$, C. Megan Urry$^1$}

\affil{$^1$Yale Center for Astronomy \& Astrophysics, Physics Department, P.O. Box 208120, New Haven, CT 06520, USA
$^2$Department of Astronomy, Faculty of Physical and Mathematical Sciences, Universidad de Concepci\'on, Casilla 160-C, Concepci\'on, Chile
$^3$Astronomy Department, Wesleyan University, Middletown, CT 06459
$^4$Department of Physics and Astronomy 3905, University of Wyoming, 1000 E. University, Laramaie, WY 82071, USA
$^5$Department of Physics, Drexel University, 3141 Chestnut Street, Philadelphia, PA 19104, USA 
$^6$Department of Astronomy and Astrophysics, and Institute for Gravitation and the Cosmos, The Pennsylvania State University, 525 Davey Lab, University Park, PA 16802
$^7$Department of Physics and Astronomy, The Johns Hopkins University, 3400 North Charles Street, Baltimore, MD 21218, USA
$^8$Department of Astronomy \& Physics, Saint Mary's University, Halifax, NS B3H 3C3, Canada
}

\begin{abstract}
SDSS J015957.64+003310.5 is an X-ray selected, $z=0.31$ AGN from the Stripe 82X survey that transitioned from a Type 1 quasar to a Type 1.9 AGN between 2000 and 2010. This is the most distant AGN, and first quasar, yet observed to have undergone such a dramatic change. We re-observed the source with the double spectrograph on the Palomar 5m telescope in July 2014 and found that the spectrum is unchanged since 2010. From fitting the optical spectra, we find that the AGN flux dropped by a factor of 6 between 2000 and 2010 while the broad H$\alpha$ emission faded and broadened. Serendipitous X-ray observations caught the source in both the bright and dim state, showing a similar 2-10 keV flux diminution as the optical while lacking signatures of obscuration. The optical and X-ray changes coincide with $g$-band magnitude variations over multiple epochs of Stripe 82 observations. We demonstrate that variable absorption, as might be expected from the simplest AGN unification paradigm, does not explain the observed photometric or spectral properties. We interpret the changing state of J0159+0033 to be caused by dimming of the AGN continuum, reducing the supply of ionizing photons available to excite gas in the immediate vicinity around the black hole. J0159+0033 provides insight into the intermittency of black hole growth in quasars, as well as an unprecedented opportunity to study quasar physics (in the bright state) and the host galaxy (in the dim state), which has been impossible to do in a single sources until now.
\end{abstract}

\section{Introduction}
Lying in the centers of galaxies, supermassive black holes grow by accreting matter that falls within their gravitational potential. Such active galactic nuclei (AGN) can be identified by signatures in their optical spectra, such as broad and/or narrow high ionization emission lines, that reveal the presence of a very deep gravitational potential or a non-stellar ionizing continuum. Spectral emission lines from gas deep within the potential well of the black hole are Doppler broadened, giving rise to the broad line region (BLR), while the narrow line region (NLR) gas, though primarily ionized by the AGN continuum, extends hundreds of parsecs beyond the accretion disk. Type 1 AGN are those systems showing both broad (full-width half-max, FWHM, $\sim$1000s km s$^{-1}$) and narrow (FWHM $\sim$ 100s km s$^{-1}$) emission lines in their optical spectra; type 2 AGN lack broad lines. Intermediate AGN types exist, including Types 1.8 and 1.9, classified by a weak, or absent, broad H$\beta$ line while retaining a broad component to H$\alpha$ \citep{osterbrock}. The simplest AGN unification models attribute differences in AGN type to the viewing angle towards an axisymmetric, parsec scale obscurer, which can block our direct view of the AGN central engine \citep{antonucci,urry}. But what happens when multiple AGN types are seen in the same source at different epochs?

Such systems are rare but not unheard of. Variable X-ray absorption has been observed in what have been dubbed ``changing-look'' AGN \citep[e.g.,][]{matt,puccetti,bianchi,risaliti,marchese}; these variations result from gas clouds moving into and out of the line-of-sight. Perhaps more striking are the AGN that have optically transitioned from a pure Type 1 state to a Type 1.8-2.0, or vice versa, with the dramatic appearance or disappearance of a strong H$\beta$ component. Within the past 40 years, such changes have been detected in only a handful of AGN, including Mrk 590 \citep{denney}, NGC 2617 \citep{shappee}, NGC 7603 \citep{tohline}, Mrk 1018 \citep{cohen}, NGC 1097 \citep{storchi-bergmann}, NGC 3065 \citep{ngc3065} and NGC 7582 \citep{aretxaga}. These types of changes might also be explained via variable absorption: in models where the obscuring material has a patchy distribution \citep[e.g.,][]{elitzur_2012}, the dynamical movement of dust clouds could, in principle, result in a change of classification. It has also been argued that changes in type are expected with variations in accretion rate \citep{elitzur} or transient events \citep[e.g., the tidal disruption of a star by a black hole;][]{eracleous}. 

\begin{figure}[ht]
{\includegraphics[scale=0.8,angle=90]{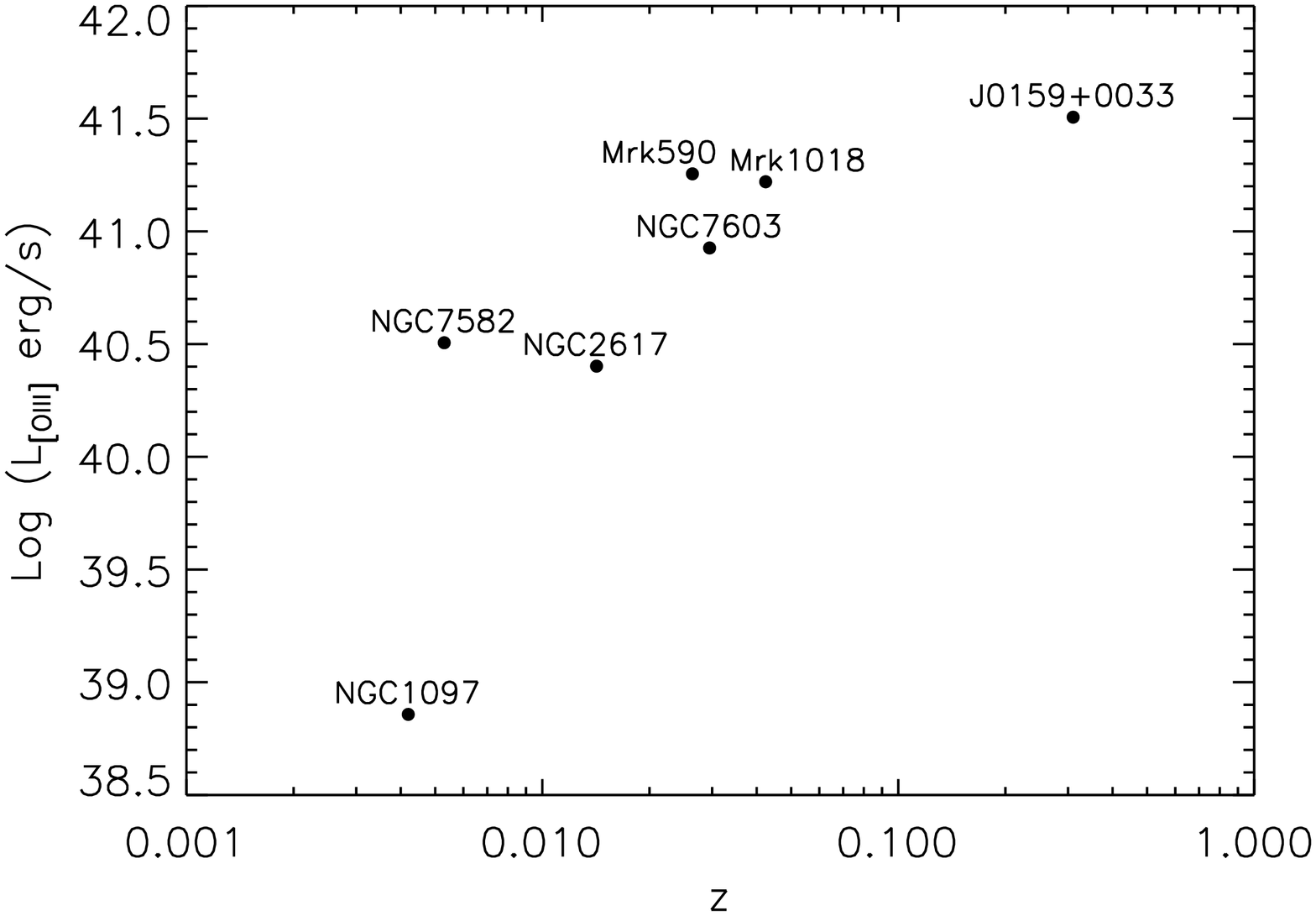}}
\caption[]{\label{o3_v_z} [OIII] luminosity \citep[proxy of intrinsic AGN luminosity,][]{heckman,me2010} as a function of redshift for changing-look AGN that have transitioned from/to a pure Type 1 state to/from a Type 1.8-2 state. With the exception of NGC 7582 which briefly transitioned from a Seyfert 2 to a Seyfert 1, the L$_{\rm [OIII]}$ values shown here reflect the intrinsic AGN power when these sources are in the bright state: as [OIII] forms in the NLR, it can take tens to hundreds of years for the dimming of the AGN continuum to propagate to the NLR and for the [OIII] emission to respond to this change. As this plot illustrates, J0159+0033 is the most distant and luminous changing-look AGN yet discovered. }
\end{figure}

Here we report the discovery of a $z=0.31$ changing-look quasar, the most distant and luminous changing AGN to date (Figure \ref{o3_v_z}). This source, SDSS J015957.64+003310.5 (hereafter J0159+0033 for brevity) is an X-ray selected AGN from the Stripe 82X survey \citep{me1,me2}, which was initially reported in the 2001 Sloan Digital Sky Survey (SDSS) Data Release 1 \citep[DR1;][]{sdss_dr1}, and was classified as a quasar in DR1, DR3, DR5 and DR7 based on its $i$-band luminosity and broad emission lines \citep{schneider_dr1,schneider_dr3,schneider_dr5,schneider_dr7}. However, a later 2010 SDSS-III \citep{sdss-iii} BOSS \citep{boss} spectrum revealed that the broad H$\beta$ component completely disappeared, yet a weak, broad H$\alpha$ component remained visible (i.e., Type 1.9 state); it was no longer included in subsequent SDSS-III quasar catalogs from DR9 and DR10 \citep{paris_dr9,paris_dr10}. We re-observed this source with the Double Spectrograph (DBSP) on Palomar in 2014 July, finding it to be in the same state as 2010. Below we discuss our analysis of the optical spectra in tandem with archival X-ray spectra and optical photometry. As we elaborate in \S\ref{discussion}, these independent lines of observational evidence taken together paint a compelling picture that variable absorption can not be the driver of the observed state change. Rather, the quasar continuum faded, providing less power to ionize the gas in the BLR. As this is the first discovery of a changing state in a quasar, this source provides a direct probe of intermittent quasar activity. Throughout, we adopt a cosmology where H$_{\rm 0}$ = 70 km/s/Mpc, $\Omega_{\rm M}$ = 0.27 and $\Lambda_{\rm 0}$ = 0.73, giving a luminosity distance of 1624 Mpc for J0159+0033.

\section{Data Analysis}
\subsection{SDSS Spectra and Photometry}
J0159+0033 was first observed spectroscopically by SDSS on 2000 November 11 through a 3$^{\prime\prime}$ diameter fiber over the wavelength range 3800-9200 \AA\ at a spectral resolution of $\simeq$2000 \citep{blanton}. The data were processed with the usual SDSS spectroscopic pipeline \citep{stoughton}. J0159+0033 was re-observed by SDSS-III BOSS on 2010 January 5 through a 2$^{\prime\prime}$ diameter fiber over a wavelength range of 3610 - 10140 \AA\ \citep{boss}. We downloaded the pipeline processed spectra for this analysis, but rather than relying on the pipeline produced fit parameters, we modeled the spectra as described in \S\ref{fitting}.

\begin{figure}[ht]
{\includegraphics[scale=0.8,angle=90]{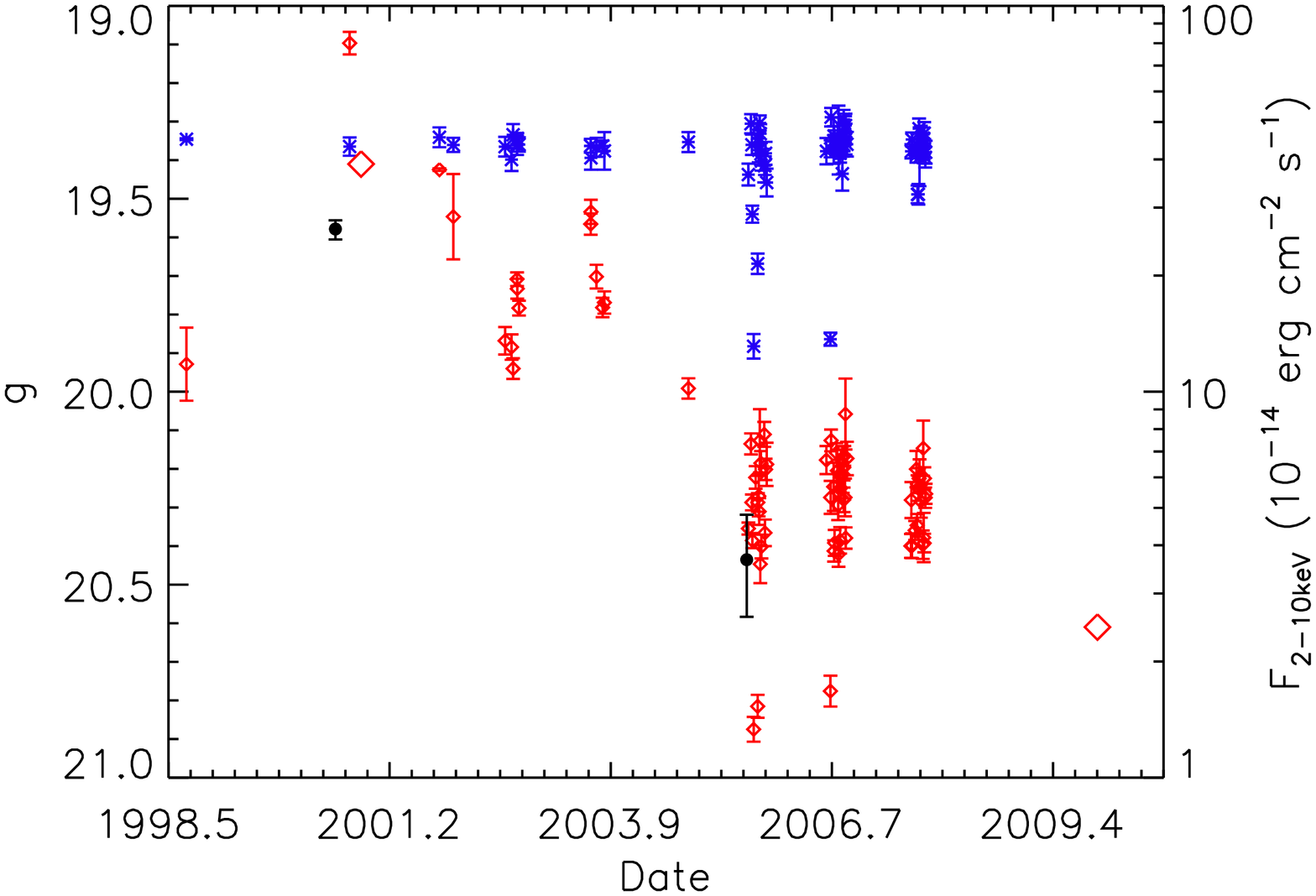}}
\caption[]{\label{mag} $g$-band PSF magnitude at various epochs for J0159+0033 (red diamonds) and a nearby reference star (blue asterisks); the larger red diamonds indicate the J0159+0033 magnitude calculated by convolving the SDSS and BOSS spectra with the $g$-band response filter. While the reference star remains at a steady brightness, the magnitude of the quasar changes, indicating that this variability is a physical result and not an effect of observing conditions. Observed hard X-ray fluxes from serendipitous {\it XMM-Newton} and {\it Chandra} observations are shown by the black circles: the optically bright and dim states are contemporaneous with the X-ray bright and dim states. }
\end{figure}

As J0159+0033 was observed in the Stripe 82 region of SDSS, which was imaged multiple times as part of a supernova survey \citep{frieman,annis}, photometric observations are available over many epochs. In Figure \ref{mag}, we plot the $g$-band PSF magnitude over the multiple observations taken in Stripe 82 (red) along with the magnitudes of a nearby (RA = 30.009, Dec=0.5474) reference star (blue). J0159+0033 increased in brightness between 1998 and 2000, after which it faded from 19.1 to 20.3 between the 2000 SDSS and 2010 BOSS observations, corresponding to a flux decrease of a factor of 3.02; the steady magnitude of the reference star indicates the change in the AGN magnitude is a physical effect and not due to observing conditions. As we show below, the non-variable emission from the host galaxy dilutes the optical variability from the changing AGN. Fortuitously, an archival {\it XMM-Newton} observation caught the source in a bright optical state while a {\it Chandra} observation detected it in the fainter optical state.

\subsection{Palomar DBSP Spectrum}
J0159+0033 was observed with the DBSP on the Palomar 200-inch Hale telescope on 2014 July 23 using a 1.5$^{\prime\prime}$ slit-width. The 2D spectrum was analyzed in {\it IRAF}\footnote{IRAF is distributed by the National Optical Astronomy Observatories, which are operated by the Association of Universities for Research in Astronomy, Inc., under cooperative agreement with the National Science Foundation.} \citep{iraf}, following the usual reduction procedures to account for bias and flat-fielding effects and a 1D spectrum was extracted from a 2.3$^{\prime\prime}$ window along the slit. The spectrum was corrected for telluric absorption features and flux calibrated using the standard star BD+28 4211. For this analysis, we used only the spectrum from the red arm of the spectrograph (where the Balmer lines are observed), limited to observed wavelengths 6000 \AA\ - 9000 \AA\ to minimize loss of sensitivity at the edges of the observed spectral range.

\subsection{\label{fitting}Optical Spectral Fitting}
Prior to fitting, all spectra were de-redshifted to the rest-frame and corrected for Galactic extinction ($E(B-V) = 0.0267$, based on the extinction quoted in the headers of the SDSS spectral files). We used the IRAF package {\it specfit} \citep{specfit} to fit the optical spectra using an array of star-formation templates generated with the program GALAXEV for stellar population ages ranging from 56 Myr to 10 Gyr (S. Charlot \& G. Bruzual 2007, private communication); a power-law component for the AGN continuum; Fe II and UV Fe templates over a range of velocities to account for blends in Fe emission lines in the broad line spectrum \citep{uvfe,feii}; and a suite of Gaussian components to model the broad and narrow emission lines. The {\it specfit} task interpolates over the templates to find the best fit via $\chi^2$ minimization for the stellar age and velocity widths (blended Fe II and UV Fe lines).

The properties of the  broad and narrow lines resulting from the fit to each spectrum are listed in Table \ref{line_fluxes}. In the 2000 SDSS spectrum, where broad H$\alpha$ and H$\beta$ are easily discernible, the narrow emission lines were fit in tandem with the broad components: the wavelengths of the broad and narrow H$\alpha$ and H$\beta$ components were tied together during the fitting. For all spectra, the flux of [OIII] 4959 \AA\ ([NII] 6548) was fixed to 33\% of the [OIII] 5007 \AA\ ([NII] 6583) flux; the [OIII] ([NII]) lines were constrained to have the same FWHMs. When fitting the SDSS and BOSS spectra, the FWHM of the (narrow) H$\beta$, [OIII] 5007 and [NII] 6583 lines were tied together: physically, we expect that these lines emanate from the narrow-line region gas which is excited by the AGN continuum, and should have the same profile. That these line complexes are well fitted by such a model indicates that a change in spectral resolution along the dispersion axis, which can affect the DBSP spectrum, is not present. However, we allowed the FWHMs to be fitted independently between [OIII] 5007 and [NII] 6583 in the DBSP spectrum to mitigate the effects of varying spectral resolution.

The stellar component was best constrained from fitting the 2010 BOSS spectrum, where continuum emission and absorption lines, such as Ca H and K transitions (at 3968.5 and 3933.7 \AA\, respectively) constrain the age of the stellar population \citep[e.g.,][]{cales}. The host galaxy emission is not expected to change within the $\sim$6.9 year (source-frame) window between observations, so when fitting the 2000 SDSS and 2014 DBSP spectra, we froze the stellar age to the best-fit found from modeling the 2010 BOSS spectrum. The 2014 DBSP spectral fit did not include an AGN power-law model as it was unconstrained and consistent with zero. Additionally, the wavelength coverage on the red side of the spectrum was not wide enough to constrain the continuum around H$\alpha$, causing the broad component to be poorly modeled. Consequently, we froze the FWHM of the broad H$\alpha$ line to the value found from modeling the 2010 BOSS spectrum. Finally, we found that the blended FeII and UV Fe components were unconstrained when fitting the 2000 SDSS spectrum (i.e., the errors on the normalizations were larger than the best-fit values), and were subsequently omitted from the final fit.

\begin{figure}[ht]
{\includegraphics[scale=0.8,angle=90]{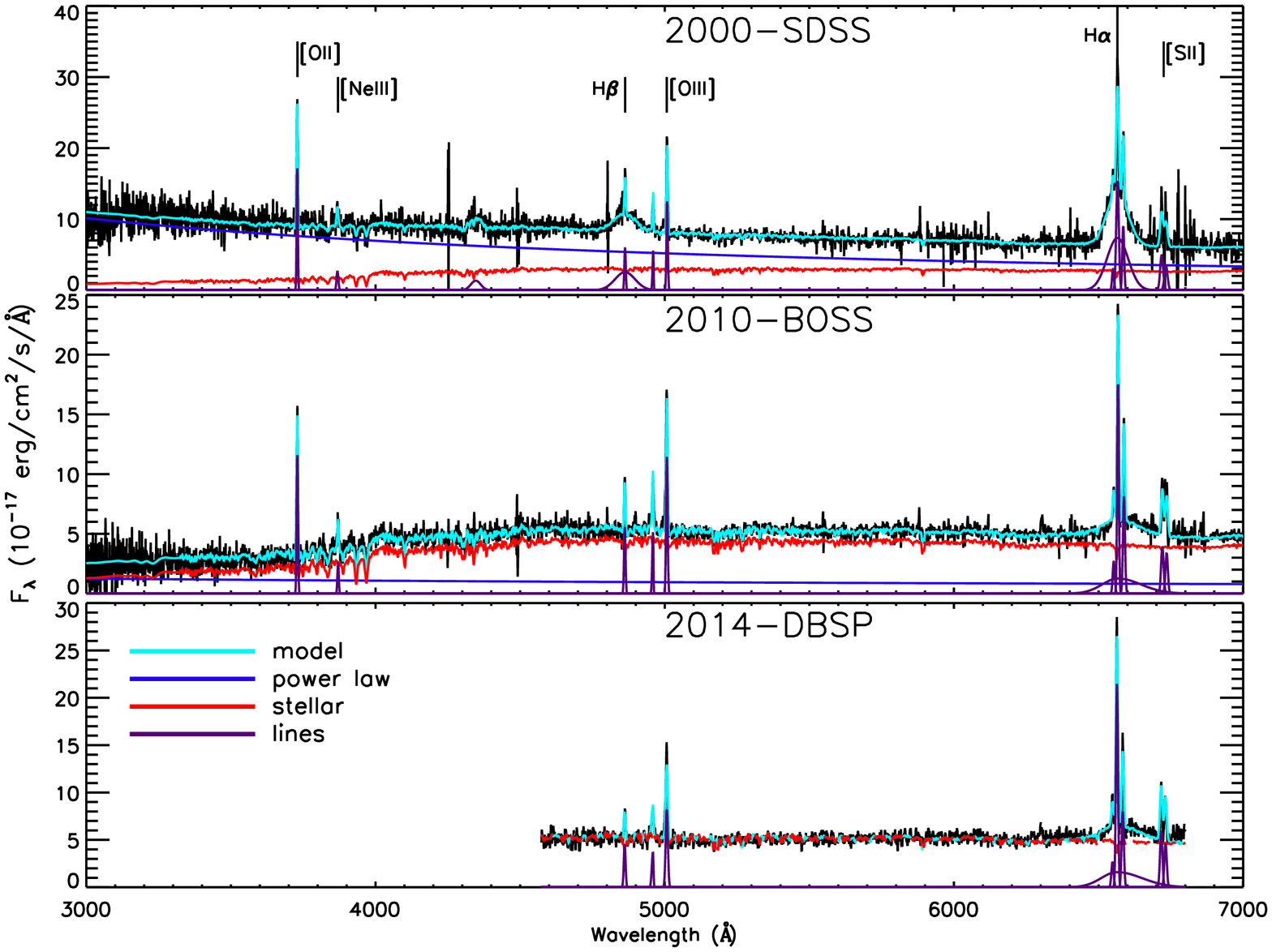}}
\caption[]{\label{specfit}Best-fit models to the optical spectra, with individual components indicated by the legend. The spectral state changed from Type 1 (broad emission lines) to Type 1.9 (only a broad component to H$\alpha$) between the years 2000 and 2010.}
\end{figure}

\begin{figure}[ht]
{\includegraphics[scale=0.8,angle=90]{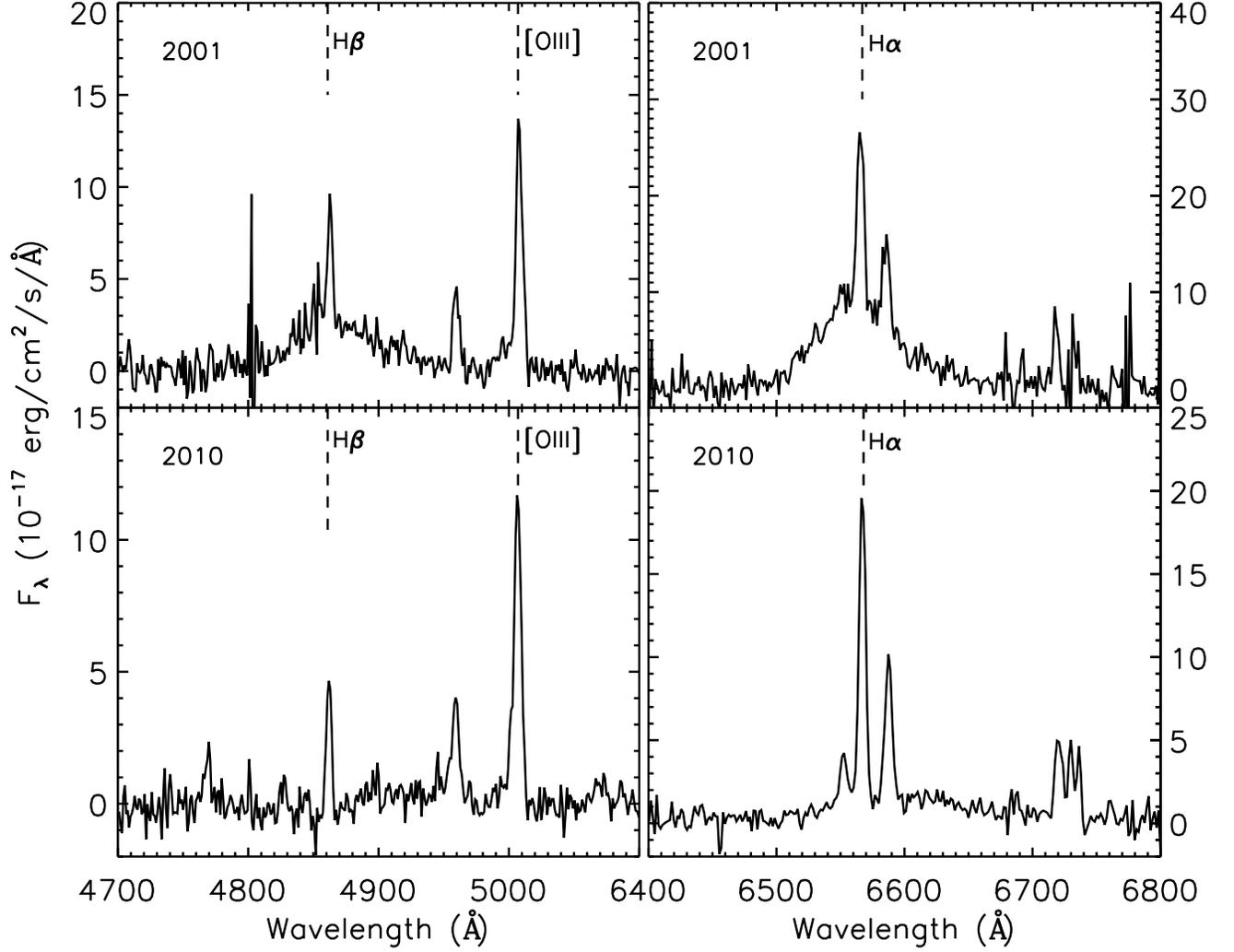}}
\caption[]{\label{ha_hb_sdss}Close-up of the H$\beta$ ({\it left}) and H$\alpha$ ({\it right}) complexes for the 2000 SDSS ({\it top}) and 2010 BOSS ({\it bottom}) spectra, where the AGN power-law continuum and galaxy components are subtracted out. The differences in the broad components between the 2 epochs are visually striking. }
\end{figure}

The best-fit models, along with individual components, for the different epochs are overplotted on the spectra in Figure \ref{specfit}, with the parameters summarized in Tables \ref{line_fluxes} and \ref{fit_params}. The most dramatic change between 2000 and 2010 is the complete disappearance of broad H$\beta$ and the near-disappearance of broad H$\alpha$, as highlighted in Figure \ref{ha_hb_sdss}. However, an asymmetric broad H$\alpha$ component exists in both the 2010 BOSS and 2014 DBSP spectra. Hence, the classification of J0159+0033 changed from a Type 1 to a Type 1.9 \citep{osterbrock} within a 9 year time-span in the observer frame ($\sim$6.9 years in source frame). The narrow emission lines remained relatively consistent among all three spectra. Narrow emission lines have been observed to vary in response to changes in the AGN ionizing continuum in other AGN, namely Mrk 590 \citep{denney} and NGC 5548 \citep{peterson}, though such changes occurred years after dimming of the central engine. As we will discuss further in Section \ref{agn_prop}, observing such variations provide direct constraints on the size of narrow line emitting region.

\begin{deluxetable}{lllllll}

\tablewidth{0pt}
\tablecaption{\label{line_fluxes}Emission Line Fits\tablenotemark{1}}
\tablehead{& \multicolumn{2}{c}{2000 (SDSS)} &  \multicolumn{2}{c}{2010 (BOSS)} &  \multicolumn{2}{c}{2014 (DBSP)} \\
& \colhead{Flux} & \colhead{FWHM} & \colhead{Flux} & \colhead{FWHM} & \colhead{Flux} & \colhead{FWHM}}

\startdata
\multicolumn{5}{c}{{\it Broad lines}}\\
H$\gamma$ & 56.9$\pm$9.7  & 2763$\pm$476 & - & - & - & - \\
H$\beta$  & 220$\pm$15     & 5043$\pm$466 & - & - & - & - \\
H$\alpha$ & 671$\pm$30    & 3917$\pm$174 & 210$\pm$18 & 5869$\pm$777\tablenotemark{2} & 282$\pm$14 & 5869\tablenotemark{3} \\
\multicolumn{5}{c}{{\it Narrow lines}}\\
{\rm [OII] 3727}   & 86$\pm$14    & 380$\pm$66  & 67.4$\pm$10.0 & 438$\pm$68  & - & -  \\
{\rm [NeIII] 3868} & 20.9$\pm$9.5  & 565$\pm$511 & 15.0$\pm$3.4  & 415$\pm$115 & - & -  \\
H$\beta$           & 35.5$\pm$6.7  & 342$\pm$42  & 35.6$\pm$5.2  & 372$\pm$40  & 34.3$\pm$3.8 & 437$\pm$56 \\
{\rm [OIII] 5007}  & 75.9$\pm$9.7 & 342\tablenotemark{4}  & 75.8$\pm$8.8 & 372\tablenotemark{4} & 78.6$\pm$6.4 & 536$\pm$46 \\
H$\alpha$          & 173$\pm$29  & 480$\pm$104 & 127$\pm$26  & 301$\pm$73  & 192$\pm$19 & 381$\pm$33 \\
{\rm [NII] 6583}   & 73.4$\pm$13.4 & 348\tablenotemark{4}  & 71.1$\pm$9.9 & 372\tablenotemark{4} &  77.4$\pm$7.6 &  412$\pm$46 \\
{\rm [SII] 6716}   & 64.7$\pm$10.8 & 547$\pm$435 & 42.4$\pm$7.9 & 452$\pm$191  & 62.8$\pm$6.5 & 442$\pm$60 \\
{\rm [SII] 6730}   & ... \tablenotemark{5} &  ...  & 35.2$\pm$6.3 & 434$\pm$61  & 48.3$\pm$6.5 &  439$\pm$75 \\

\enddata
\tablenotetext{1}{Observed flux in units of 10$^{-17}$ erg cm$^{-2}$ s$^{-1}$ and FWHM in units of km s$^{-1}$.}
\tablenotetext{2}{Broad component of H$\alpha$ in the 2010 BOSS spectrum has a skew (asymmetry) of 1.45$\pm$0.16.}
\tablenotetext{3}{FWHM of the broad component of H$\alpha$ frozen to the best-fit value found from fitting the 2010 BOSS spectrum; the fitted skew in the DBSP spectrum is 1.55$\pm$0.10.}
\tablenotetext{4}{FWHM of [OIII] 5007 and [NII] 6583 tied to H$\beta$ during spectral fitting.}
\tablenotetext{5}{FWHM of [SII] 6730 was unconstrained during the spectral fitting of the 2000 SDSS spectrum, so we do not report the best-fit values.}
\end{deluxetable}

\begin{deluxetable}{llll}

\tablewidth{0pt}
\tablecaption{\label{fit_params}Galaxy and AGN Parameters}
\tablehead{&  \colhead{2000 (SDSS)} & \colhead{2010 (BOSS)} & \colhead{2014 (DBSP)} }

\startdata
\multicolumn{3}{c}{{\it AGN power-law}} \\
Normalization (1000 \AA\ ) & 42.9$\pm$0.5     & 2.3$\pm$0.1   & -  \\
$\alpha_\lambda$            & 1.31$\pm$0.01 & 0.5 $\pm$0.1 & - \\
\multicolumn{3}{c}{{\it Stellar Component}} \\
Normalization & 19.4 $\pm$0.2  & 28.9$\pm$0.8  &  34.5$\pm$0.1\\
age (Gyr)     & 1.64\tablenotemark{1}  & 1.64$\pm$0.03 & 1.64\tablenotemark{1}\\

\enddata
\tablenotetext{1}{Stellar age was frozen to the value found from fitting the 2010 BOSS spectrum, where the host galaxy properties could be better constrained. The age of the stellar population does not vary within the observed time scales reported here.}
\end{deluxetable}

\subsection{X-ray Spectra}
Both the {\it Chandra} and {\it XMM-Newton} archival X-ray observations of J0159+0033 were serendipitous: the AGN happened to be in the field-of-view of observations targeting other sources. The {\it XMM-Newton} data from 2000 (ObsID: 0101640201) were reduced with {\it SAS} tasks to filter for flaring, create good time intervals and extract a spectrum from each of the three detectors (PN, MOS1, MOS2). The net exposure time was 10 ks, with 2375 counts detected among the three detectors. Similarly, the {\it Chandra} data from 2005 (ObsID: 5777) were processed with {\it CIAO} tools {\it chandra\_repro} to create a cleaned events file from which a spectrum was extracted with {\it specextract}. Though the net exposure time was twice that of {\it XMM-Newton}, only 148 counts were detected.

Although {\it Chandra} has better spatial resolution than {\it XMM-Newton}, we extracted the {\it Chandra} spectrum using the same source region size as {\it XMM-Newton} (i.e., 30$^{\prime\prime}$) to ensure that any possible changes in the X-ray spectra over time are due to physical variations and not unresolved extended emission in the {\it XMM-Newton} extraction. In both cases, the background was subtracted from an annulus with an inner radius of 45$^{\prime\prime}$ and outer radius of 90$^{\prime\prime}$. The {\it XMM-Newton} spectra were grouped with 20 counts per bin while the {\it Chandra} spectrum was grouped with 5 counts per bin. The spectra were fitted independently in XSpec with an absorbed power-law model, where one absorption component was frozen to the Galactic value and another absorption component was set at the redshift of the source and left free. We used the $\chi^2$ statistic for {\it XMM-Newton} and C-statistic for {\it Chandra}. Neither spectrum has features that warrant more complex modeling.

\begin{figure}[ht]
\begin{centering}
{\includegraphics[scale=0.5,angle=270]{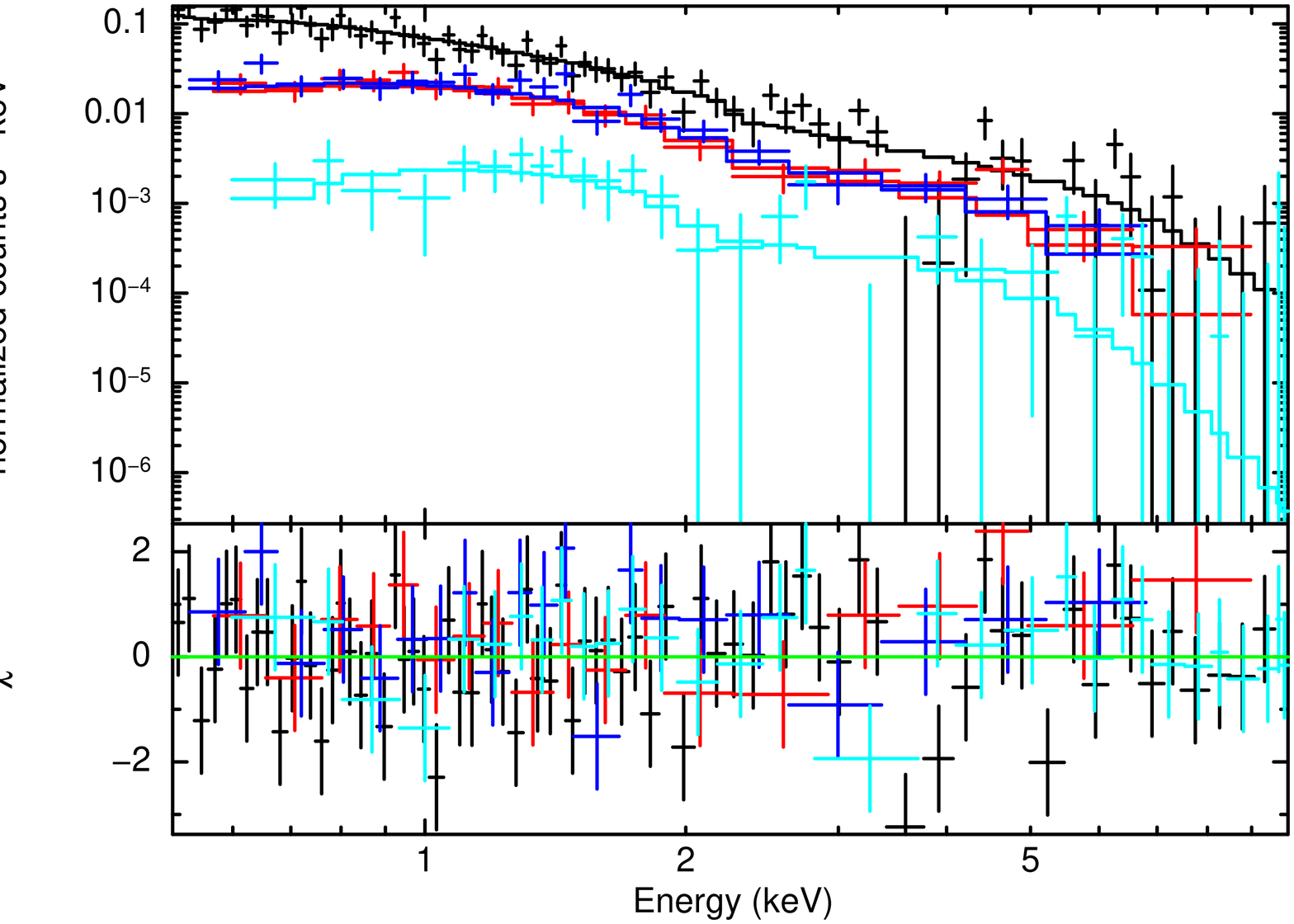}}
\caption[]{\label{x-ray} Observed-frame X-ray spectra (data points) with absorbed power-law model fit (solid lines) overplotted for the 2000 {\it XMM-Newton} observation (black: PN detector, red: MOS1 detector, blue: MOS2 detector) and 2005 {\it Chandra} observation (cyan). Serendipitously, the {\it XMM-Newton} observation was coincident with the bright optical state while the {\it Chandra} spectrum was observed after the source dimmed (see Figure \ref{mag}). The source dimmed by a factor of $\sim$10 between observations with no change in absorption (see text for details).}
\end{centering}
\end{figure}

Figure \ref{x-ray} shows the {\it XMM-Newton} and {\it Chandra} spectra of J0159+0033, with the absorbed power law model overplotted. The photon indices measured from the {\it XMM-Newton} and {\it Chandra} spectra are consistent within the error bars (2.13$\pm$0.1 and 1.81$^{+0.53}_{-0.49}$, respectively). In both cases, the fitted absorption at the redshift of the source was consistent with zero, with a 90\% confidence level upper limit of N$_{\rm H}<$ 3$\times10^{20}$ and 5$\times10^{21}$ cm$^{-2}$ between the {\it XMM-Newton} and {\it Chandra} observations, though the higher upper limit from the {\it Chandra} spectrum is likely a statistical effect due to low signal-to-noise in the spectrum. There is one dramatic change between epochs: the observed 2-10 keV flux dropped from 26.4$^{+1.4}_{-1.6}\times 10^{-14}$ erg cm$^{-2}$ s$^{-1}$ to 3.7$\pm1.1\times 10^{-14}$ erg cm$^{-2}$ s$^{-1}$, a factor of 7.2. As the spectra had similar photon indices when fit independently,\footnote{The flattening of the spectrum, though not significant within the errors, could reflect an intrinsic change in the spectrum as steeper photon indices are associated with higher luminosity sources.} we fit the spectra simultaneously, tying together the photon indices and including a fixed component of absorption due to our Galaxy, allowing the normalization to vary to account for flux changes. With that analysis, we find that the observed 2-10 keV flux dropped by a factor of $\sim$12.3 and is not associated with a change in intrinsic absorption. We also note that in addition to the absence of absorption above that from our Galaxy, other signatures of X-ray absorption, such as an Fe K$\alpha$ emission feature at 6.4 keV \citep[from X-ray photons reprocessed within a circumnuclear obscuring medium;][]{krolik} or strong spectral curvature between 2-6 keV, are markedly absent. To change the flux by variable absorption alone would have produced easily visible spectral features.

\section{Discussion}
\subsection{\label{agn_prop}AGN properties}
From the measured FWHM of the broad H$\beta$ emission line component and the monochromatic 5100 \AA\ luminosity calculated from the AGN power-law fit to the broad-line 2000 SDSS spectrum ($\lambda$L$_{\rm 5100}$ = 1.09$\pm$0.01 $\times 10^{44}$ erg s$^{-1}$), we can estimate the mass of the central black hole \citep{mbh_hb}:
\begin{equation}
M_{\rm BH} = 10^{6.91} \left[\frac{{\rm FWHM(H\beta)}}{1000 {\rm \; km \; s^{-1}}}\right]^2 \left[\frac{\lambda L_{\rm 5100}}{10^{44} {\rm \; erg \; s^{-1}}}\right]^{0.5}\;{\rm M}_\odot,
\end{equation}

\noindent finding a value of 2.2$\pm0.4 \times10^{8}$ M$_{\sun}$. 

Alternatively, we can estimate the black hole mass from the broad H$\alpha$ component from both the 2000 SDSS and 2010 BOSS spectra using the relation of \citet{mbh_ha}:
\begin{equation}
M_{\rm BH} = 9.7\times10^{6} \left[\frac{{\rm FWHM(H\alpha)}}{1000 {\rm \; km \; s^{-1}}}\right]^{2.06} \left[\frac{\lambda L_{\rm 5100}}{10^{44} {\rm \; erg \; s^{-1}}}\right]^{0.519}\;{\rm M}_\odot,
\end{equation}

\noindent where we find a black hole mass that is similar between the two epochs (1.7$\pm0.1\times10^{8}$ M$_{\sun}$ and 1.6$\pm0.4\times10^{8}$ M$_{\sun}$ in 2000 and 2010, respectively), and is consistent with the value from the H$\beta$ calculation. Though the H$\alpha$ FWHM changed drastically between the years 2000 and 2010, it was compensated by a decrease in AGN continuum luminosity, where $\lambda$L$_{\rm 5100}$ measured from the 2010 BOSS spectrum is 2.0$\pm$0.1 $\times 10^{43}$ erg s$^{-1}$. As we discuss below, this has implications for the amount of circumnuclear gas ionized by the central engine.

We estimate the bolometric AGN luminosity (L$_{\rm bol}$) to be 8.8$\times10^{44}$ erg s$^{-1}$, using $\lambda$L$_{\rm 5100}$ \citep[see][for a discussion of the reliability of various bolometric luminosities]{richards} and a conversion factor of 8.1 to convert from monochromatic to bolometric luminosity \citep{runnoe}. With the Eddington luminosity defined as L$_{\rm Edd} = 1.3\times10^{38}$ M$_{\rm BH}$ \citep{frank}, we find an Eddington parameter ($\lambda_{\rm Edd}$ = L$_{\rm bol}$/L$_{\rm Edd}$) of 0.04. If we estimate L$_{\rm bol}$ using $\lambda$L$_{\rm 5100}$ from the spectral fit when the source was in the Type 1.9 state (with the caveat that the utility of $\lambda$L$_{\rm 5100}$ as a proxy for L$_{\rm bol}$ and the bolometric correction are based on studies of Type 1 AGN), L$_{\rm bol}$ becomes 1.6$\times10^{44}$ erg s$^{-1}$ and $\lambda_{\rm Edd}$  drops to 0.007.

Additionally, we estimate the characteristic radius of the broad-line region in 2000 using the $R-L$ relation calibrated by \citet{bentz}.
More specifically, this is the characteristic radius corresponding to the H$\beta$ reverberation time lag. The equation reads

\begin{equation}
\log\left[\frac{R_{\rm BLR}}{1 \; {\rm lt-days}}\right] = 1.527 + 0.533 \, \log\left[\frac{\lambda L_{\rm 5100}}{10^{44} {\rm \; erg \; s^{-1}}}\right],
\label{eq:RBLR}
\end{equation}

\noindent which implies $R_{\rm BLR}\sim$35 light days.  The gas in the BLR that contributes to the broad-line flux, however, extends to radii that are a few time larger than this value. This picture is supported by the following observational results. (a) Reverberation mapping of the H$\beta$ line shows that the response of the line flux extends to a maximum lag that is about a factor of 2 longer than the radius of peak response. Moreover lower-order Balmer lines have a longer peak lag than higher order Balmer lines (for example, the peak lag of H$\alpha$ is about a factor of 2 longer than that of H$\beta$). Good examples of this behavior can be seen in the velocity-delay maps presented by \citet{grier13} and \citet{bentz10}. (b) The time lag of the response of the optical \ion{Fe}{2} line complex is 2-3 times longer than the lag of the H$\beta$ line \citep{barth13} and this \ion{Fe}{2}-emitting gas is also a source of Balmer lines \citep[e.g.,][]{baldwin04}. Therefore, the BLR gas that is of interest here is distributed out to distances {\it at least} 3 times larger than the distance corresponding to the H$\beta$ lag. The issue of the extent of the BLR gas is relevant to our discussion of the reddening hypothesis for the observed variability in \S\ref{sec:reddening}.

The NLR is larger and may extend dozens to thousands of parsecs depending on the luminosity and redshift of a quasar \citep[e.g.,][]{bennert,yonehara,oh}, where the limiting size of the NLR may be set by the gas available for ionization \citep{hainline1,hainline2}. Thus, the response of this gas to the changing ionization source will be delayed compared with the BLR and X-ray emission. With future monitoring campaigns, decline of the narrow line strengths might be observed, should the NLR be sufficiently compact enough for this change to be observationally feasible, which would allow the size of the NLR to be directly constrained. Indeed, the gas in the NLRs of Mrk 590 and NGC 5548 has responded to the decrease in the AGN ionizing continuum \citep{denney,peterson}, indicating that variations in the NLR are sometimes observeable in a single object.

\subsection{\label{discussion}Reddening or Dimming of the AGN Continuum?\label{sec:reddening}}
The observed flux diminution and disappearance of the broad H$\beta$ and (most of the) H$\alpha$ lines can result from either variable absorption blocking the line-of-sight to the broad-line, but not the narrow-line, region, or a change in the AGN continuum ionizing the BLR. To test the first possibility, we applied reddening to the AGN model components  from the 2000 SDSS spectrum (continuum and broad lines) to determine if we can reproduce the observed 2010 BOSS spectrum.\footnote{The spectral state shows no sign of change between the 2010 BOSS and 2014 DBSP spectra, so we use the 2010 BOSS spectrum for this analysis as it has a wider wavelength coverage.} We use the {\it specfit}-provided \citet{cardelli} extinction curve, with $R_V$ set to 3.1 \citep{osterbrock2}; we note that different choices of $R_V$ do not affect the extinction curve in the optical bandpass \citep{glikman}. To fit the 2010 spectrum, we froze the AGN power-law continuum and broad emission line model components to the values from the 2000 SDSS spectral fit (see Table \ref{line_fluxes}), allowing $E(B-V)$ and the narrow emission line parameters to be free.

\begin{figure}[ht]
{\includegraphics[scale=0.8,angle=90]{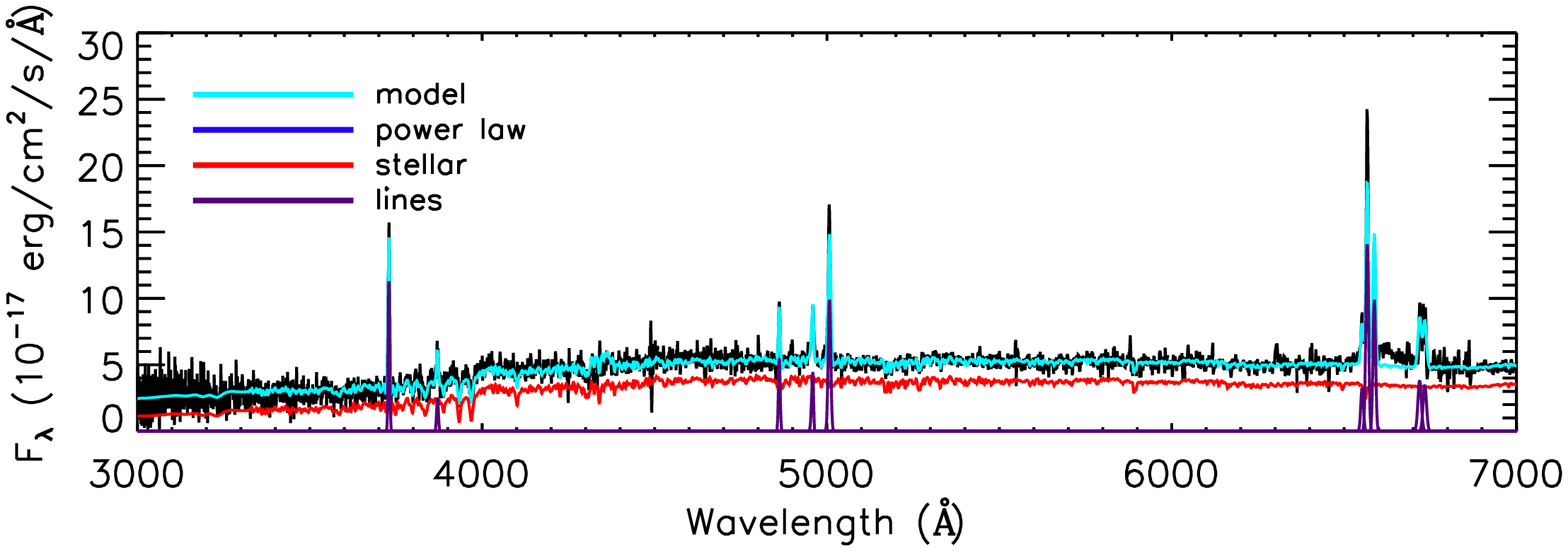}}
\caption[]{\label{red_fit} 2010 BOSS spectrum fitted with a model where the broad emission lines and AGN power-law were frozen to their best-fit values from the 2000 spectrum, but were attenuated by a foreground screen of dust; the narrow emission lines, stellar component and $E(B-V)$ were the only parameters allowed to be free. This toy model can reproduce the general shape of the 2010 spectrum, except for the H$\alpha$ complex (see Figure \ref{h_alpha} and Section 3.2 of the text for details). }
\end{figure}

\begin{figure}[ht]
{\includegraphics[scale=0.4,angle=90]{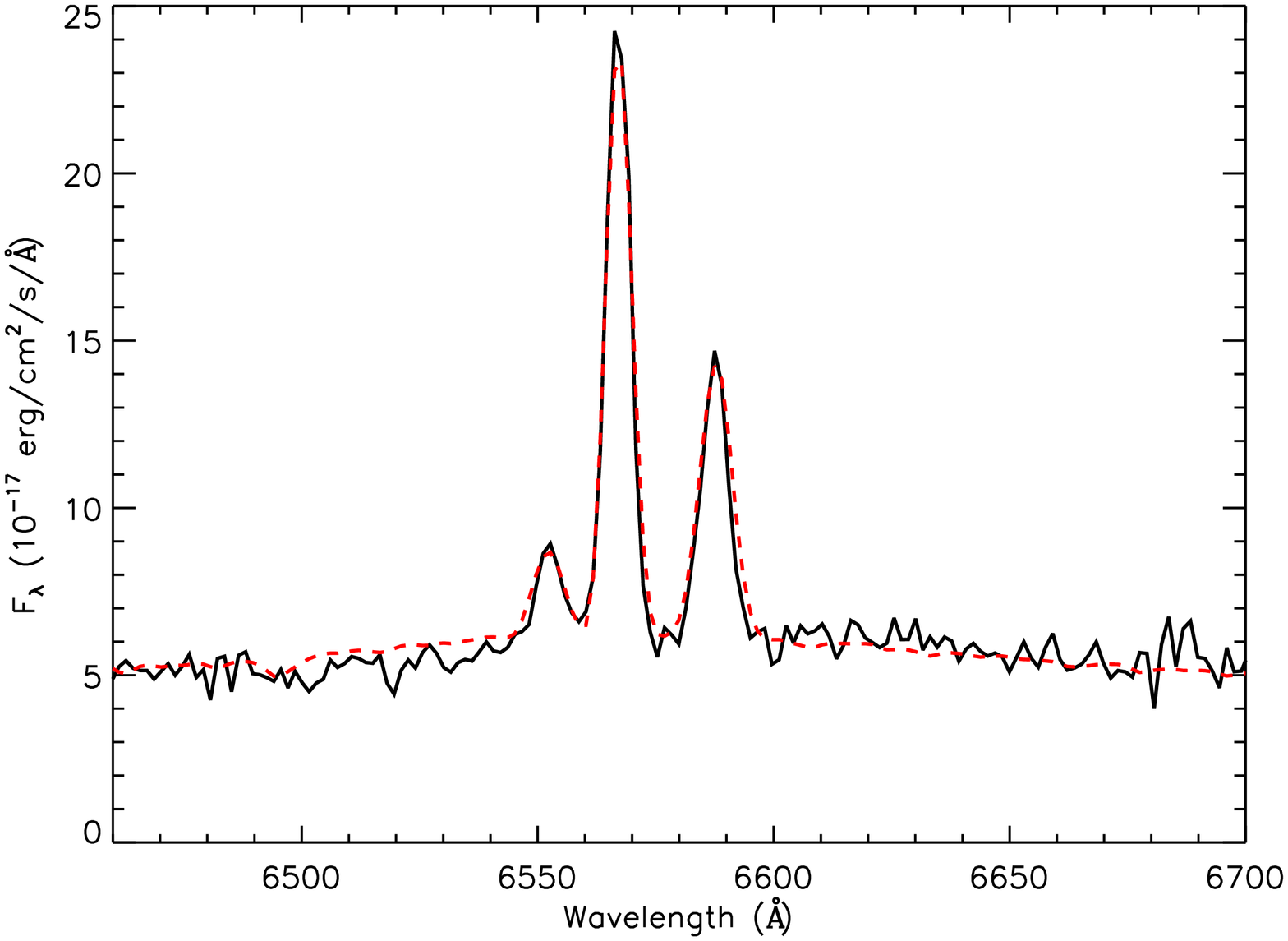}}
{\includegraphics[scale=0.4,angle=90]{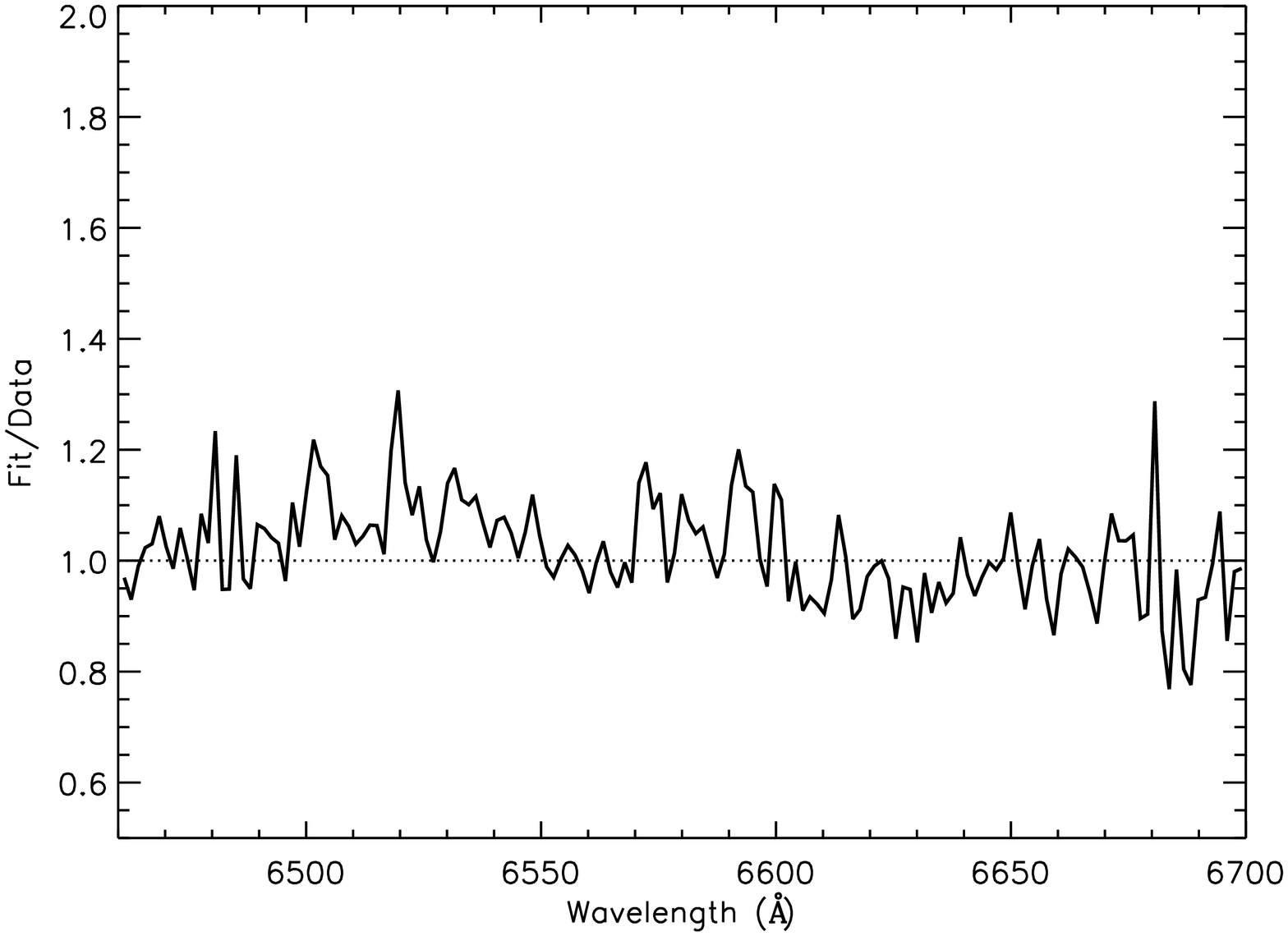}}
{\includegraphics[scale=0.4,angle=90]{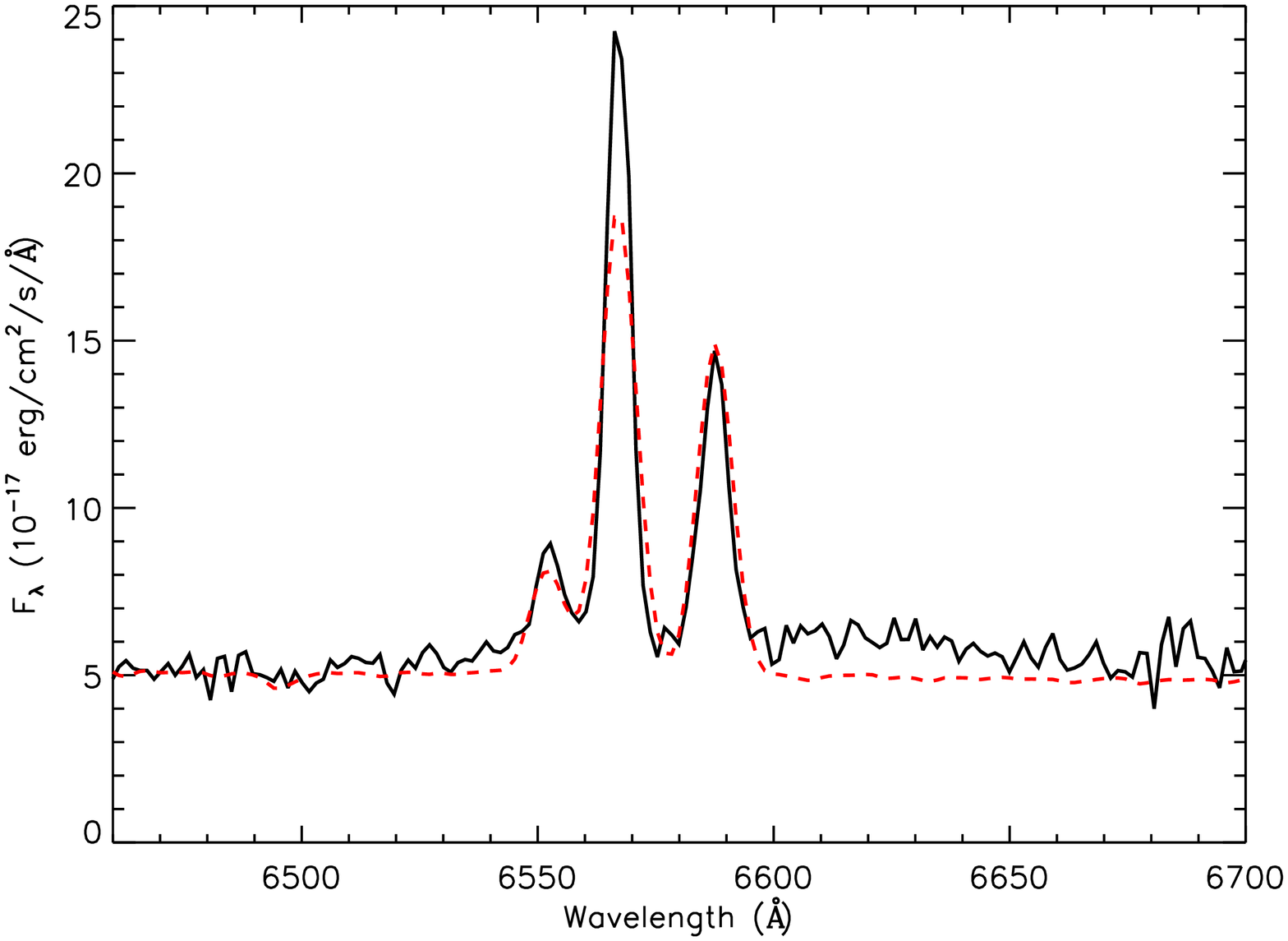}}
{\includegraphics[scale=0.4,angle=90]{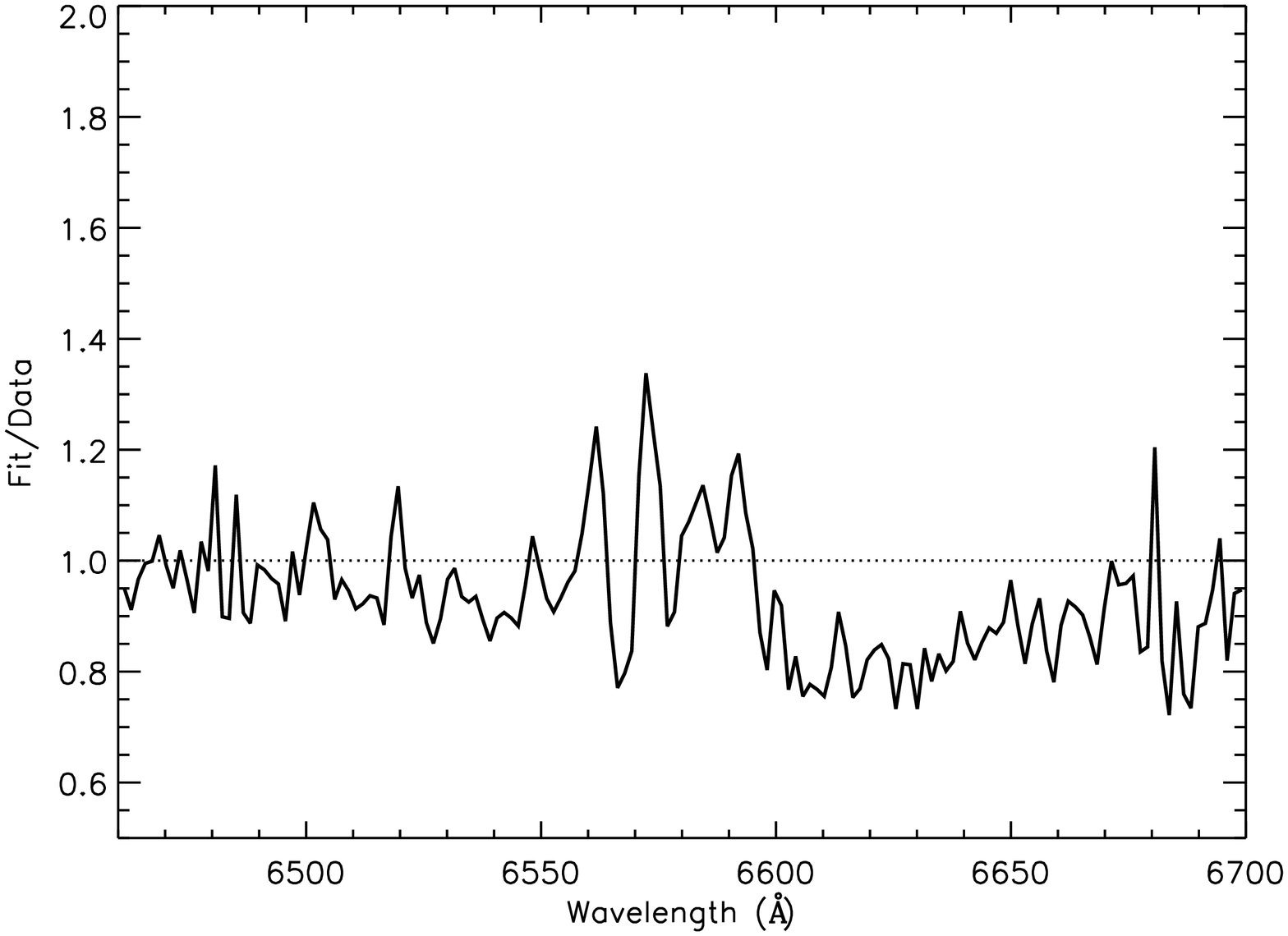}}
\caption[]{\label{h_alpha} Model fit ({\it left}) and ratio of model-to-data ({\it right}) for the H$\alpha$ complex of the 2010 BOSS spectrum by fitting the spectrum with a broad H$\alpha$ component and AGN continuum ({\it top}) or applying reddening to the broad emission lines and AGN continuum model values derived from fitting the 2000 SDSS spectrum ({\it bottom}): as shown in the bottom panels, reddening alone under-predicts the strength of the broad H$\alpha$ component, resulting in a poor fit to the broad and narrow H$\alpha$ emission lines and the [NII] 6548,6583 lines.  This poor fit indicates that variable absorption is not responsible for the observed spectral change between 2000 and 2010. }
\end{figure}

With a fitted $E(B-V)$ value of 0.42, the general shape, and many of the narrow emission lines, from the 2010 BOSS spectrum can be reproduced (Figure \ref{red_fit}). However, as shown in Figure \ref{h_alpha}, the H$\alpha$ complex is not well accommodated by this model, especially when compared with our previous fit to the 2010 BOSS spectrum. Specifically, the observed broad H$\alpha$ line is substantially stronger than would be predicted in a scenario where reddening alone is responsible for the spectral state change. Additionally, since the broad H$\alpha$ component is not a free parameter in this fit and not appropriately accomodated by this absorption model, the narrow emission lines superimposed on this feature are subsequently poorly modeled.

Furthermore, we can estimate the crossing time for an intervening object orbiting outside the BLR on a Keplerian orbit as
\begin{equation}
t_{\rm cross} = 0.07  \left[\frac{r_{\rm orb}}{1 {\rm lt-day}}\right]^{3/2} M_{8}^{-1/2}\; {\rm arcsin}\left[\frac{r_{\rm src}}{r_{\rm orb}}\right] \; {\rm yr},
\label{eq:tcross}
\end{equation}
\noindent where r$_{\rm orb}$ is the orbital radius of the foreground object, M$_{8}$ is the black hole mass in units of 10$^{8}$M$_{\sun}$, and r$_{\rm src}$ is the true size of the BLR. The above equation was derived by assuming that the orbiting object is on a circular, Keplerian orbit around the black hole and by evaluating the time needed for this object to travel the length of an arc that corresponds to the projected size of the BLR, i.e. $t_{\rm cross}=\Delta\phi/\omega_K$, where $\Delta\phi$ is the angular length of the arc and $\omega_K$ is the Keplerian frequency. In the most conservative case, i.e. the case that minimizes the crossing time, the intervening object should be at $r_{\rm orb} \ge R_{\rm BLR} =  35\;$lt-days (see equation~\ref{eq:RBLR}), so that it can intercept a substantial fraction of the broad Balmer line flux. Even in the extreme limit where we assume $r_{\rm orb} = R_{\rm BLR}$  and M$_{8} = 1.7$, the crossing time of such an obscurer would be $\sim$10 years. More realistically, the intervening object should be placed at 
$r_{\rm orb} \ge 3\,R_{\rm BLR} = 105\;$lt-days, so that it can cover a substantial portion of the gas in the BLR, specifically gas at low velocities that contributes flux to the cores of the line profile, in which case we obtain $t_{\rm cross}=20\;$years.

Not only are these time scales too long to explain the observations shown here, but it is not obvious whether intervening material with the necessary physical properties required to obscure the entire continuum region and BLR could exist at such radii. For instance, one possibility for such an obscuring medium located at these distances is the obscuring torus that lies at the outskirts of the BLR, beyond the dust sublimation radius. Based on near-IR reverberation mapping campaigns \citep[e.g.,][]{suganuma06, koshida14}, this sublimation radius is 4--5 times larger than the optical emitting BLR: movement of this material into the line of sight to cover the BLR would be set by the same, long time scales as the general obscurer just discussed. Another obscuration scenario is the possibility that the dimming of the continuum facilitates the formation, not just transport, of dust that can cover the BLR. However, assuming a gas density of 10$^5$ cm$^{-3}$, appropriate for the torus \citep[e.g.][]{nenkova02}, the dust formation time scale is $\sim$10$^3$~years \citep[see equation~(8) of][]{draine09}, which is much too long for this scenario to be viable.

Additionally, variable absorption fails to explain other observables: broadening and skewness (asymmetry) of the H$\alpha$ line in the latter epoch; the lack of absorption signatures in the X-ray spectra when the X-ray emission diminished; and a similar drop in observed X-ray and optical fluxes (a factor of 7.2 vs. 5.5, respectively), since absorption should affect the latter much more than the former if the same source of obscuration is responsible for suppressing both the X-ray and optical emission. To produce this latter effect via obscuration, a dust cloud of low optical depth would need to block the line-of-sight to the BLR while a gas cloud of much higher optical depth would have to attenuate emission from the X-ray emitting corona, where the column densities of both clouds would have to be fine tuned so that the optical and X-ray fluxes drop by the same amount. 

An alternate hypothesis is that the AGN continuum faded, diminishing the power available to ionize the gas around the black hole. Indeed, the fact that the fitted H$\alpha$ emission profile broadened between 2000 and 2010, resulting in the same black hole masses being measured using the H$\alpha$ and $\lambda$L$_{\rm 5100}$ values obtained from each epoch, suggests that such a dimming took place: as the continuum faded, fewer ionizing photons were available to excite the BLR gas, causing the subsequent emission to be concentrated over a smaller range of radii,\footnote{Using $\lambda$L$_{\rm 5100}$ measured from the 2010 BOSS spectrum, we estimate that R$_{\rm BLR}$ shrank to $\sim$14 light days.} and we are thereby sampling regions closer to the black hole where the velocities are higher. Furthermore, the gas in the BLR orbits the black hole with a Keplerian velocity, such that the FWHM of the emitted broad lines scales as  $R_{\rm BLR}^{-1/2}$. Based on reverberation mapping studies, $R_{\rm BLR} \sim L^{1/2}$ \citep[e.g.,][]{davidson,krolik78,bentz06,bentz09,bentz}, where $L$ is the AGN continuum luminosity, so that the FWHM scales as $L^{-1/4}$. Between 2000 and 2010, the broad-component to H$\alpha$ decreased by a factor of 0.67, which is almost identical to $(\lambda$L$_{\rm 5100,2000}$/$\lambda$L$_{\rm 5100,2010}$)$^{-1/4}$, i.e., 0.65, providing additional evidence that we have a direct view of the BLR.
 
The change in the broad emission lines may be explained by the paradigm proposed by \citet{elitzur}, who demonstrated that a decrease in accretion rate is associated with a decrease in bolometric luminosity powering the BLR, which leads to an evolution in spectral type. They find a trend between AGN type and Eddington rate and L$_{\rm bol}$/M$_{\rm BH}^{2/3}$, which systematically decreases along the sequence from Type 1 $\rightarrow$ 1.2/1.5 $\rightarrow$ 1.8/1.9  $\rightarrow$ 2.0. Our calculated  L$_{\rm bol}$/M$_{\rm BH}^{2/3}$ values (39.5 and 38.7 dex between the Type 1 and Type 1.9 stage, respectively) are consistent with the values they calculate for Type 1 and Type 1.8 AGN from the \citet{stern} sample (mean values of 39.39 and 38.93 dex, respectively), which is used to inform their modeling.

To assess whether a rapid change in accretion rate is a plausible mechanism for the observed dimming of the continuum, we estimate the inflow time scale of gas in the inner parts of the accretion disk, where the UV continuum primarily originates. The inflow time scale is the time for a parcel of gas in the disk to move radially from a given radius to the center. It is also the time scale on which the luminosity of the continuum emitted from the disk can fluctuate as a result of abrupt changes in the accretion rate. Assuming the Shakura-Sunyaev accretion disk model \citep{shakura} and using the formulae therein, we find that radiation pressure dominates over gas pressure in the inner portion of the disk of J0159+0033, given the black hole mass and Eddington parameter determined above. Accordingly, we cast the inflow time as
\begin{equation}
t_{\rm infl} = 31\; \left[\frac{\alpha}{0.1}\right]^{-1}\,  \left[\frac{\lambda_{\rm Edd}}{0.03}\right]^{-2}\,  \left[\frac{\eta}{0.1}\right]^{2}\,  \left[\frac{r}{10 \, r_{\rm g}}\right]^{7/2} 
\left[\frac{M_{8}}{1.7}\right],
\label{eq:tinflow}
\end{equation}
where $\alpha$ is the Shakura-Sunyaev ``viscosity'' parameter, $\eta$ is the efficiency of converting potential energy to radiation, and $r_{\rm g}$ is the gravitational radius ($GM/c^2$). We have scaled the Eddington parameter and the black hole mass to the values appropriate for the 2000 SDSS spectrum J0159+0033 and adopted commonly assumed values for the other parameters, including a radius of $10\,r_{\rm g}$ for the UV emitting region of the disk. At face value, the inflow time appears too long to be compatible with the observed variability time scale, however, numerical simulations of black hole accretion flows \citep[e.g.][]{krolik05} indicate that the infall speed in the inner disk is at least several times higher than the value given by the Shakura-Sunyaev model. Thus, the inflow time could plausibly be of the order of a few years, which {\it is} compatible with the observed variability time scale. The inflow time scale can also become shorter than the estimate of equation~(\ref{eq:tinflow}), if other mechanisms of angular momentum loss are invoked, such as large scale waves in the inner disk or hydromagnetic winds. It is noteworthy that the optical continuum is produced in a region of the disk that is several times larger than the UV-emitting region \citep[see illustration in Figure 25 of][]{frank} and for which the inflow time is an order of magnitude longer. Nonetheless, the observed variability of the optical continuum is probably driven by reprocessing of UV or X-ray light, as discussed by \citet{krolik91} and \citet{cackett} and can be as rapid as the variability in those higher-frequency bands. It may be possible to test the accretion rate fluctuation scenario, if we can observe the recovery of the luminosity of J0159+0033 to its earlier level. According to equation~(\ref{eq:tinflow}), the recovery time scale should be considerably longer than the dimming time scale since the Eddington parameter is an order of magnitude lower in the dim state.

An alternative explanation for abrupt luminosity changes may be found in different processes that operate on much shorter time scales, such as the thermal time of the inner accretion disk or its hot corona that is thought to be the source of the X-ray emission. In fact, abrupt changes in the size of X-ray emitting corona have been observed in other AGN, namely PHL~1092 \citep{miniutti} and Mrk~335 \citep{gallo}, though these sources do not show variations in their optical spectra.

A future spectropolarimetric observation of J0159+0033 may discriminate between the obscuration and intrinsic dimming scenarios: if a broad H$\beta$ component is viewed in polarized light, this would indicate that obscuration could explain the spectral state change \citep[e.g.,][]{tran},  though a lack of this signature would not necessarily indicate that hidden broad emission lines do not exist \citep[see][for a review]{antonucci2012}.

\section{The Story of J0159+0033}
The different lines of observational evidence taken together suggest the following history for J0159+0033 over the past decade and a half. Between 1998 and 2000, the source brightened optically and was observed spectroscopically to contain broad and narrow emission lines.  The object stays in a bright state for several years before fading by a magnitude compared to its brightness in 1998. $\lambda_{\rm Edd}$ drops from  $\sim$4\% to $\sim$0.7\% Eddington between the bright and dim stage. During this time, the fraction of ionized gas in the BLR decreases in response to the dimming of the central engine. The optical dimming of J0159+0033 coincides with a decrease in X-ray flux, which is expected as the hot corona in which the X-rays form is relatively compact and near the accretion disk, thereby responding to changes in the AGN continuum on timescale much faster than the extended NLR. The fact that the narrow lines continue to be visible at least 8 years after the source dimmed, and at strengths consistent with those measured in the 2000 SDSS spectrum, demonstrate that the black hole was not dormant before the observed increase in brightness in 2000: the NLR is not compact enough to have ``turned on'' between 1998 and 2000, otherwise we would have observed significant decreases in the fluxes of NLR emitting gas between 2000 and 2014. Continued monitoring of the source will reveal if the source transitions to a pure Type 2 state, as predicted by \citet{elitzur}, and when the narrow lines weaken in response to the decrease in energy output from the central engine. This time scale will provide an estimate of the NLR size, should the NLR be sufficiently compact for this process to be observable within the next few decades.

Extreme changes observed in the optical spectra of AGN are quite rare, and though this has been detected before, J0159+0033 is much more distant and luminous than previously reported changing-look AGN. It is the first true quasar discovered to have undergone such a dramatic fading. It is interesting to note that most reported cases of spectral transition states (the exceptions being NGC 1097 and NGC 3065 which were LINERs and NGC 7582 which was a Sy2) are from a pure Type 1 state to an intermediate type, or vice versa; that is, a broad component to H$\alpha$ and/or H$\beta$ never completely disappeared. In some cases, the classification of AGN as an intermediate type is consistent with reddening along the line of sight to the broad line region \citep{goodrich1,goodrich2,goodrich3}, while this explanation has been ruled out for other objects \citep{goodrich2,goodrich3,trippe}. In fact, in a recent systematic analysis of a large number of intermediate type AGN at L$_{\rm bol}<10^{44}$ erg s$^{-1}$, \citet{stern} demonstrated that these systems are largely unobscured.  \citet{elitzur} used the \citet{stern} sample to model the dependence of spectral type with accretion state to support the notion of an evolutionary pathway. In J0159+0033, the observed decrease in L$_{\rm bol}$/M$_{bh}^{2/3}$ between the Type 1 and Type 1.9 state are consistent with the \citet{elitzur} evolutionary paradigm, but only as long as the accretion rate can fluctuate fast enough to account for the observed continuum variability.

Observations of these changing-look AGN represent a valuable opportunity to directly compare black hole properties (when in the AGN-dominated state) to host galaxy properties (when the AGN fades), a comparison that is typically impossible for luminous quasars since they swamp out the light from their host. Future monitoring will determine whether the quasar is shutting off or if the spectral states reported here represent intermittent quasar activity, providing the first template where a quasar lifetime can be accurately estimated \citep{Martini}.  Indeed, a possible avenue to discover a population of sources similar to J0159+0033, that is, changing-look quasars beyond the immediate Universe is provided by the Time Domian Spectroscopic Survey (TDSS, Morganson et al. 2014; carried out within SDSS~IV), which will obtain new spectra of SDSS quasars that have changed their brightness by more than 0.8 magnitudes in the past decade. We have shown that one such object exists, which provides at least one data point to estimate the duty cycle of quasar activity after such statistical searches are completed. In addition, observations of these changing-look AGN demonstrate that AGN unification is an over-simplifying interpretation of AGN spectral types, and that in some instances, AGN optical classification may represent the ability of the central engine to ionize gas in its vicinity, and possibly, an evolutionary stage as the gas powering accretion ultimately disappears forever into the black hole.

\acknowledgements
We thank the referee for a thorough reading of this manuscript and very insightful comments. We thank Nic Ross, Eilat Glikman, Joe Hennawi, Ryan Hickox, Julian Krolik, and Ari Laor for useful discussions. ADM was partially supported by NASA ADAP award NNX12AE38G and EPSCoR award NNX11AM18A and by NSF award 1211112. 

Funding for the SDSS and SDSS-II has been provided by the Alfred P. Sloan Foundation, the Participating Institutions, the National Science Foundation, the U.S. Department of Energy, the National Aeronautics and Space Administration, the Japanese Monbukagakusho, the Max Planck Society, and the Higher Education Funding Council for England. The SDSS Web Site is http://www.sdss.org/.

The SDSS is managed by the Astrophysical Research Consortium for the Participating Institutions. The Participating Institutions are the American Museum of Natural History, Astrophysical Institute Potsdam, University of Basel, University of Cambridge, Case Western Reserve University, University of Chicago, Drexel University, Fermilab, the Institute for Advanced Study, the Japan Participation Group, Johns Hopkins University, the Joint Institute for Nuclear Astrophysics, the Kavli Institute for Particle Astrophysics and Cosmology, the Korean Scientist Group, the Chinese Academy of Sciences (LAMOST), Los Alamos National Laboratory, the Max-Planck-Institute for Astronomy (MPIA), the Max-Planck-Institute for Astrophysics (MPA), New Mexico State University, Ohio State University, University of Pittsburgh, University of Portsmouth, Princeton University, the United States Naval Observatory, and the University of Washington.

Funding for SDSS-III has been provided by the Alfred P. Sloan Foundation, the Participating Institutions, the National Science Foundation, and the U.S. Department of Energy Office of Science. The SDSS-III web site is http://www.sdss3.org/.

SDSS-III is managed by the Astrophysical Research Consortium for the Participating Institutions of the SDSS-III Collaboration including the University of Arizona, the Brazilian Participation Group, Brookhaven National Laboratory, Carnegie Mellon University, University of Florida, the French Participation Group, the German Participation Group, Harvard University, the Instituto de Astrofisica de Canarias, the Michigan State/Notre Dame/JINA Participation Group, Johns Hopkins University, Lawrence Berkeley National Laboratory, Max Planck Institute for Astrophysics, Max Planck Institute for Extraterrestrial Physics, New Mexico State University, New York University, Ohio State University, Pennsylvania State University, University of Portsmouth, Princeton University, the Spanish Participation Group, University of Tokyo, University of Utah, Vanderbilt University, University of Virginia, University of Washington, and Yale University.

\end{document}